\documentclass[12pt]{iopart}
\usepackage{graphicx}
\usepackage{multirow}
\usepackage{epsfig}
\usepackage{dcolumn}
\usepackage{color}
\usepackage{multicol}
\usepackage{bm}
\usepackage{iopams}  

\begin{document}

\title{Contact parameters in two dimensions for general three-body systems}
\author{F F Bellotti$^{1,2,3}$, T Frederico$^{1}$, M T Yamashita$^4$, D V Fedorov$^2$, A S Jensen$^2$ and N T Zinner$^2$}
\address{$^1$ Instituto Tecnol\'{o}gico de Aeron\'autica, 12228-900, S\~ao Jos\'e dos Campos, SP, Brazil \\
$^2$ Department of Physics and Astronomy, Aarhus University, DK-8000 Aarhus C, Denmark \\
$^3 $Instituto de Fomento e Coordena\c{c}{\~a}o Industrial, 12228-901, S{\~a}o Jos{\'e} dos Campos, SP, Brazil \\
$^4$ Instituto de F{\'i}sica Te{\'o}rica, UNESP - Univ Estadual Paulista, C.P. 
70532-2, CEP 01156-970, S{\~a}o Paulo, SP, Brazil} 

\begin{abstract}
We study the two dimensional three-body problem in the general case of
three distinguishable particles interacting through zero-range
potentials. The Faddeev decomposition is used to write the
momentum-space wave function.  We show that the large-momentum
asymptotic spectator function has the same functional form as derived 
previously for three identical particles. We derive analytic relations 
between the three different Faddeev components for three distinguishable
particles.  We investigate the one-body momentum distributions both
analytically and numerically and analyze the tail of the distributions
to obtain two- and three-body contact parameters.  We specialize from
the general cases to examples of two identical, interacting or
non-interacting, particles.  We find that the two-body contact
parameter is not a universal constant in the general case and show
that the universality is recovered when a subsystem is composed of two
identical non-interacting particles.  We also show that the three-body
contact parameter is negligible in the case of one non-interacting
subsystem compared to the situation where all subsystem are bound.  As
example, we present results for mixtures of Lithium with two Cesium or
two Potassium atoms, which are systems of current experimental
interest.
\end{abstract}
\maketitle

\section{Introduction}
The surprising and unexpected predictions from quantum mechanics
have been challenging our classical intuition for a century. 
Since then, efforts in both theoretical and experimental fields have increasingly
aimed at a better understanding of quantum systems. In particular, 
experiments with cold atomic gases \cite{bloch2008} are an interesting way of building and 
probing quantum systems: properties of atomic condensates near absolute 
zero temperature are governed by pure quantum effects. An interesting 
example of an unexpected quantum mechanical prediction that was experimentally 
confirmed in cold atomic gases is the so-called Efimov effect, 
which predicts that three identical bosons interacting through short range 
potentials present infinitely many bound states, where the energies 
between states are geometrically spaced. This effect was predicted by Efimov 
in 1970 \cite{efi70} and was experimentally verified in cold atomic gases 
experiments in 2006 \cite{kraemer2006}. The Efimov effect happens for three dimensional (3D) 
systems, while the quantum theory predicts that the same two dimensional (2D) 
system presents only two bound states and no Efimov effect \cite{bru79,nie97,nie01,adh88,adh93}. 
The theoretical difference arising 
from changing system dimension could most likely also be verified in cold atomic gases 
experiments in the near future since experimentalists are already able to 
change dimensionality of such systems and build experiments in effectively
one (1D) or two (2D) spatial dimensions.

Another important theoretical prediction, that was recently reported in 
\cite{tan2008}, is the emergence of a parameter in the study of two-component 
Fermi gases, the two-body contact parameter, $C_2$, which connects universal 
relations between two-body correlations and many-body properties
\cite{tan2008,werner2012a,braaten2008,braaten2010,langmack2012,combescot2009,schakel2010,braaten2011,castin2011,werner2012b,valiente2011}. 
This parameter can most easily be defined by considering the 
single-particle momentum distribution of the system, $n(\bm q)$. In the 
limit where $q\to\infty$, one finds
\begin{equation}
\lim_{q\to\infty}q^4 n(q)=C_2.
\end{equation}
As mentioned above, this parameter appears in a number of universal relations
for both few- and many-body properties.
These relations also hold for bosons and were confirmed in cold atomic gases 
experiments for two-component Fermi gases \cite{stewart2010,kuhnle2010} and for bosons \cite{wild2012}. One way to 
determine this parameter is to find the coefficient in the leading order asymptotic behavior 
of the one-body momentum density of few-body systems. The next 
order in this expansion defines the three-body contact parameter, $C_3$, 
but since the Pauli principle suppresses the short-range correlations 
for two-component Fermi gases, the three-body parameter is only important 
for bosons \cite{combescot2009,werner2012b}. In a gas of identical 
bosons in 3D the Efimov effect occurs and one finds a momentum distribution 
of the form 
\begin{equation}
n(q)\to \frac{C_2}{q^4}+C_3\frac{\sin(s_0\ln(q)+\delta)}{q^5}\,\textrm{for}\,q\to\infty,
\end{equation}
where $s_0$ is the Efimov parameter \cite{efi70} and $\delta$ is constant \cite{castin2011}.

The two- and three- body contact parameters were studied for a 
3D system of three identical bosons in \cite{castin2011,braaten2011,pricoupenko2011} 
and for 
mixed-species systems in \cite{yam13}. These results show that the 
influence of non-equal masses in three-body systems goes beyond 
changing the contact parameters values. In 3D the sub-leading term, which 
defines $C_3$, changes to a different functional form when the masses
are not equal. In view 
of that, one may ask what changes could we get when dealing with mixed 
species systems in 2D. It is known that mixed-species systems have a 
richer energy spectrum in 2D comparing with symmetric mass 
systems \cite{lim1980,pricoupenko2010,bel13b}. The study of the momentum 
distribution for three identical bosons in 2D was reported 
in \cite{bel13a} where the two-body contact parameter is found to be a 
universal constant, in the sense that $\frac{C_2}{E_3}$ is the same 
for both three-body bound states of energy $E_3$. The leading 
order term of the momentum distribution at large momenta 
has the same inverse quartic form in 1D \cite{olshanii2003,barth2011}, 2D, and 3D. 
This can be derived on general grounds and is intuitively connected to 
the behavior of the free propagator for the particles \cite{pricoupenko2011,valiente2012}.
However, 
the three-body contact parameter and the functional form of the 
sub-leading term were showed to present a very different 
behavior in 2D as compared to 3D \cite{bel13a} although no analytical results
have been presented to estimate the value of the contact parameter.

In this paper we study cylindrically symmetric 
three-body bound states of 2D systems composed 
of three distinguishable particles with attractive short-range 
interactions. We derive analytic expressions for $C_2$ and $C_3$ and 
numerically obtain the one-body momentum density to verify our 
results. Unlike 3D systems, the sub-leading order in the asymptotic 
momentum density presents the same functional form for both equal 
masses and mixed-species systems. We find that $C_2$ no longer shows 
universal behavior in the general case but do show that the universality 
is recovered in at least one special case of two identical non-interacting 
particles. We also extend our asymptotic formulas to the full range of momenta 
and use it to give an analytic expression for $C_2$ for the ground state.

The paper is organized as follows. The formalism and the quantities that 
appear in the work are properly shown and defined in Sec.~\ref{formal}. The analytic 
formulas to the asymptotic spectator function are discussed in Sec.~\ref{spec} 
and the one-body large momentum behaviors  
are derived in Sec.~\ref{momentum}. The numeric results with an appropriate discussion 
is presented in Sec.~\ref{contact}. Discussion, conclusions, and an outlook 
are given in Sec.~\ref{discussion}.

\section{Formalism}\label{formal}
We consider the two dimensional (2D) problem of three interacting
particles of masses $m_A,m_B,m_C$ which are pairwise bound with
energies $E_{AB},E_{AC},E_{BC}$.  The interaction is assumed to be
described as attractive zero-range potentials and the resulting
$s-$wave three-body bound state of energy $-E_3$ is fully determined
by these six parameters: three two-body energies and three masses.  
We shall only investigate bound states and we therefore let $E_3>0$ denote the
absolute value of the binding energy.  We use the Faddeev decomposition to write the
momentum space wave function as \cite{bel11} 
\begin{equation}
\Psi\left(\mathbf{q}_\alpha,\mathbf{p}_\alpha\right)=\frac{f_{\alpha}\left(q_\alpha\right)+f_{\beta}\left(\left| \mathbf{p}_\alpha- \frac{m_\beta}{m_\beta+m_\gamma}\mathbf{q}_\alpha\right|  \right)+f_{\gamma}\left(\left| \mathbf{p}_\alpha+ \frac{m_\gamma}{m_\beta+m_\gamma}\mathbf{q}_\alpha\right| \right)}{E_{3}+\frac{q_\alpha^{2}}{2m_{\beta \gamma,\alpha}}+\frac{p_\alpha^{2}}{2m_{\beta \gamma}}} , 
\label{eq.01}
\end{equation} 
where $\mathbf{q}_\alpha,\mathbf{p}_\alpha$ are the Jacobi momenta of
particle $\alpha$, and $m_{\beta \gamma,\alpha}=
m_\alpha(m_\beta+m_\gamma)/(m_\alpha+m_\beta+m_\gamma)$ and $m_{\beta
  \gamma}= (m_\beta+m_\gamma)/(m_\beta+m_\gamma)$ are the reduced
masses.  We use here $(\alpha,\beta,\gamma)$ as cyclic permutations of
$(A,B,C)$.  The wave function is written in the Jacobi coordinate
system related to the momentum, $\mathbf{q}_\alpha$, of particle
$\alpha$ relative to the center of mass of the $\beta-\gamma$
subsystem.

The three spectator functions, $f_{\alpha,\beta,\gamma}(\mathbf{q})$,
or the momentum space Faddeev components, obey three coupled
homogeneous integral equations for any given bound state \cite{bel11}
\begin{eqnarray}
f_{\alpha}\left( \mathbf{q}\right)  = \tau_\alpha(q,E3)
&\int d^{2}k\left( \frac{f_{\beta}\left(k\right) }{E_{3}+\frac{q^{2}}{2 m_{\alpha \gamma}}+\frac{k^{2}}{2 m_{\beta \gamma}}+\frac{1}{m_\gamma}\mathbf{k}\cdot \mathbf{q}}\right.&\nonumber\\
&\left.+\frac{f_{\gamma}\left( k\right) }{E_{3}+\frac{q^{2}}{2m_{\alpha \beta }}+\frac{k^{2}}{2m_{\beta \gamma}}+\frac{1
}{m_\beta}\mathbf{k}\cdot \mathbf{q}}\right) , \label{eq.02}&
\end{eqnarray}
where
\begin{equation}
\tau_\alpha(q,E3)=\left[ 4\pi m_{\beta \gamma}\ln \left( 
\sqrt{\frac{\frac{q^{2}}{2m_{\beta \gamma,\alpha} }+E_{3}}{E_{\beta\gamma}}}
\right) \right] ^{-1}.
\label{eq.A02}
\end{equation}
The equations of motion defining the spectator functions are seen to
be invariant under the following scaling relations:
(i) multiply all energies, that is $E_3$ as well as all $E_{\beta \gamma}$
by the same constant $t_c$; (ii) multiply all masses by a constant factor $s_c$.
Then all momenta ($q_{\alpha},p_{\alpha}$) should be multiplied by
$\sqrt{t_c s_c}$. This means we can choose both a unit of energy, say $E_2$, and a unit of
mass, say $m_2$, while using the momenta in the unit of $\sqrt{m_2 E_2}$.
In other words, after all calculations are done with $E_2=1$ and $m_2=1$,
we have to multiply all energies by $E_2$, all masses by $m_2$, and all
momenta by $\sqrt{m_2 E_2}$.
The one-body momentum density of particle $\alpha$ is defined by
$n(q_\alpha)=\int{d^2 p_\alpha
  |\Psi(\mathbf{q}_\alpha,\mathbf{p}_\alpha)|^2}$, where
$\Psi(\mathbf{q}_\alpha,\mathbf{p}_\alpha)$ is given in
(\ref{eq.01}).  We use the normalization where $\int{d^2
  q_\alpha\;n(q_\alpha)}=1$.  Following the procedure in
\cite{bel13a,yam13,castin2011}, we group the nine terms in $\int{d^2
  p_\alpha |\Psi(\mathbf{q}_\alpha,\mathbf{p}_\alpha)|^2}$ into four
components with distinctly different integrand structure.  The
one-body momentum density is expressed as a sum of four terms, that is
$n(q_\alpha)=\sum_{i=1}^4{n_i(q_\alpha)}$.

A general system of three distinguishable particles, presents three
distinct one-body momentum density distributions corresponding to the
different particles.  The four terms for particle $\alpha$ can be
expressed as
\begin{eqnarray}
&n_1(q_\alpha)=\left|f_{\alpha}\left(q_\alpha\right)\right|^2  \int{d^2\;p \frac{1}{\left(E_{3}+\frac{q_\alpha^{2}}{2m_{\beta \gamma,\alpha}}+\frac{p^{2}}{2m_{\beta \gamma}}\right)^2}} = \frac{2 \pi m_{\beta \gamma} \left|f_{\alpha}\left(q_\alpha\right)\right|^2}{E_3+\frac{q_\alpha^{2}}{2m_{\beta \gamma,\alpha}}}, 
\label{eq.03a}&
\end{eqnarray}

\begin{eqnarray}
n_2(q_\alpha)= &\int{d^2\;k\frac{\left|f_{\beta}(k)\right|^2}{\left(E_{3}+\frac{q_\alpha^{2}}{2m_{\alpha \gamma}}+\frac{k^{2}}{2m_{\beta \gamma}}+\frac{\mathbf{k}\cdot \mathbf{q_\alpha}}{m_\gamma}\right)^2}} &\nonumber\\
&+\int{d^2\;k\frac{\left|f_{\gamma}(k)\right|^2}{\left(E_{3}+\frac{q_\alpha^{2}}{2m_{\alpha \beta}}+\frac{k^{2}}{2m_{\beta \gamma}}-\frac{\mathbf{k}\cdot \mathbf{q_\alpha}}{m_\beta}\right)^2}},&
\label{eq.03b}
\end{eqnarray}

\begin{eqnarray}
n_3(q_\alpha)= &2 f_{\alpha}\left(q_\alpha\right) \left[\int{d^2\;k\frac{f_{\beta}(k)}{\left(E_{3}+\frac{q_\alpha^{2}}{2m_{\alpha \gamma}}+\frac{k^{2}}{2m_{\beta \gamma}}+\frac{\mathbf{k}\cdot \mathbf{q_\alpha}}{m_\gamma}\right)^2}}\right.&\nonumber\\ 
&\left.+ 
\int{d^2\;k\frac{f_{\gamma}(k)}{\left(E_{3}+\frac{q_\alpha^{2}}{2m_{\alpha \beta}}+\frac{k^{2}}{2m_{\beta \gamma}}-\frac{\mathbf{k}\cdot \mathbf{q_\alpha}}{m_\beta}\right)^2}} \right],& 
\label{eq.03c}
\end{eqnarray}
\begin{eqnarray}
n_4(q_\alpha)= & \int{d^2\;k\frac{f_{\beta}(k)f_{\gamma}(|\mathbf{k+q_\alpha}|)}{\left(E_{3}+\frac{q_\alpha^{2}}{2m_{\alpha \gamma}}+\frac{k^{2}}{2m_{\beta \gamma}}+\frac{\mathbf{k}\cdot \mathbf{q_\alpha}}{m_\gamma}\right)^2}} \nonumber\\
&+
 \int{d^2\;k\frac{f_{\gamma}(k)f_{\beta}(|\mathbf{k+q_\alpha}|)}{\left(E_{3}+\frac{q_\alpha^{2}}{2m_{\alpha \beta}}+\frac{k^{2}}{2m_{\gamma \beta}}+\frac{\mathbf{k}\cdot \mathbf{q_\alpha}}{m_\beta}\right)^2}}, &
\label{eq.03d}
\end{eqnarray}
where the integration variable originating from (\ref{eq.01}) are
properly redefined to simplify the arguments of the spectator
functions in the integrands.  Only $n_4$ is then left with an angular
dependence through the spectator functions.  We emphasize that the
distributions for the other particles can be obtained by cyclic
permutations of $(\alpha,\beta,\gamma)$ in these expressions.

\section{Spectator functions}\label{spec}
The spectator functions are the key ingredients. They can be
characterized by their behavior in small and large momentum limits.  We
are first of all interested in large momenta, but we shall as well
extract the behavior for small momenta. Hopefully these pieces can be
put together in a coherent structure.

\subsection{Large-momentum behavior}
For three identical bosons all spectator functions are equal, and the
large-momentum behavior was previously found to be \cite{bel13a}
\begin{equation}
\lim_{q \to \infty} f(q) \to  \Gamma_0 \frac{\ln q}{q^2} \ ,
\label{eq.A00}
\end{equation}
where the constant $\Gamma_0$ depends on which excited state we focus
on.  
Corrections, $\delta f(q)$, to (\ref{eq.A00}) must vanish faster
than $\ln(q)/q^2$ for $q\to\infty$, i.e. $\delta f(q) q^2/\ln(q)\to 0$.
Henceforth, we will refer to (\ref{eq.A00}) as the large-momentum 
leading order behaviour of the spectator function.

For three distinct particles, we have three different spectator
functions.  However, their large-momentum asymptotic behavior all
remain identical, when all three two-body subsystems are bound, except
for individual proportionality factors.  To prove this we carry out
the angular integrals in (\ref{eq.02}), which immediately gives
\begin{eqnarray}
&f_{\alpha}\left( \mathbf{q}\right)  = 2\pi \tau_\alpha(q,E_3) &\nonumber\\
&\times\left[ \int_0^\infty {dk \frac{k f_{\beta}\left( k \right) }{\left(E_{3}+\frac{q^{2}}{2 m_{\alpha \gamma}}+\frac{k^{2}}{2 m_{\beta \gamma}}\right)\sqrt{1-\frac{k^2 q^2/m_\gamma^2} {\left(E_{3}+\frac{q^{2}}{2 m_{\alpha \gamma}}+\frac{k^{2}}{2 m_{\beta \gamma}}\right)^2}}}}+ \right. & \nonumber\\
& \left. + \int_0^\infty{ dk\frac{k f_{\gamma}\left(k\right)}{\left(E_{3}+\frac{q^{2}}{2m_{\alpha \beta}} +\frac{k^{2}}{2m_{\beta \gamma}}\right)\sqrt{1-\frac{k^2 q^2/m_\beta^2} {\left(E_{3}+\frac{q^{2}}{2m_{\alpha \beta }}+\frac{k^{2}}{2m_{\beta \gamma}}\right)^2}}}}\right] .& \label{eq.A01}
\end{eqnarray}
The two terms in (\ref{eq.A01}) have the same form, and one can be
obtained from the other by interchange of $\beta$ and $\gamma$.  It
therefore suffices to calculate the first integral in (\ref{eq.A01}).

The contribution for large $q$ can in principle be collected from
$k$-values ranging from zero to infinity.  To separate small and large
$k$-contributions we divide the integration into two intervals, that
is from zero to a large ($q$-independent) momentum
$\Lambda\gg\sqrt{E_3}$, and from $\Lambda$ to infinity. Thus
\begin{eqnarray}
&f_{\alpha}\left( \mathbf{q}\right)  = \tau_\alpha(q,E_3)  \left[ \int_0^\Lambda {dk \frac{k f_{\beta}\left(k\right)}{\left(E_{3}+\frac{q^{2}}{2 m_{\alpha \gamma}}+\frac{k^{2}}{2 m_{\beta \gamma}}\right)\sqrt{1-\frac{k^2 q^2/m_\gamma^2} {\left(E_{3}+\frac{q^{2}}{2 m_{\alpha \gamma}}+\frac{k^{2}}{2 m_{\beta \gamma}}\right)^2}}}}+ \right. & \nonumber\\
& \left. + \int_\Lambda^\infty {dk \frac{k f_{\beta}\left( k\right) }{\left(\frac{q^{2}}{2 m_{\alpha \gamma}}+\frac{k^{2}}{2 m_{\beta \gamma}}\right)\sqrt{1-\frac{k^2 q^2/m_\gamma^2} {\left(E_{3}+\frac{q^{2}}{2 m_{\alpha \gamma}}+\frac{k^{2}}{2 m_{\beta \gamma}}\right)^2}}}}+...\right] ,& \label{eq.A03}
\end{eqnarray}
where the dots indicate that the second term in (\ref{eq.A01}) should
be added. For $q\to\infty$ the first term, $f_{\alpha,1}$, on the
right-hand-side of (\ref{eq.A03}) goes to zero as
\begin{equation}
\lim_{q \to \infty} f_{\alpha,1}\left( q\right)  \to  \frac{m_{\alpha \gamma}/m_{\beta \gamma}}{q^2 \ln(q)}\int_0^\Lambda {dk \frac{k f_{\beta}\left(k\right) }{\sqrt{1-\frac{k^2 q^2/m_\gamma^2} {\left(E_{3}+\frac{q^{2}}{2 m_{\alpha \gamma}}+\frac{k^{2}}{2 m_{\beta \gamma}}\right)^2}}}} \ ,
\label{eq.A04}
\end{equation} 
where we used that $\tau_\alpha(q,E_3) \to \left[ 2 m_{\beta \gamma}\ln
  q \right] ^{-1}$, and that both $E_3$ and $\frac{k^{2}}{2 m_{\beta
    \gamma}}$ are much smaller than $\frac{q^{2}}{2 m_{\alpha
    \gamma}}$.
The integral in (\ref{eq.A04}) is finite and only weakly $q-$dependent
for large $q \gg \Lambda$.
The asymptotic spectator function in (\ref{eq.A00}) can be inserted in
the second term, $f_{\alpha,2}$, on the left-hand-side of
(\ref{eq.A03}), because we are in the asymptotic limit where
$k>\Lambda$.  For $q\to\infty$ we then get
\begin{eqnarray} 
&\lim_{q \to \infty} f_{\alpha,2}\left(q\right)  \to&\nonumber\\ & \frac{\Gamma_\beta}{2 m_{\beta \gamma} \ln q} \int_\Lambda^\infty {dk \frac{ \ln k} {k \left(\frac{q^{2}}{2 m_{\alpha \gamma}}+\frac{k^{2}}{2 m_{\beta \gamma}}\right)\sqrt{1-\frac{k^2 q^2/m_\gamma^2} {\left(E_{3}+\frac{q^{2}}{2 m_{\alpha \gamma}}+\frac{k^{2}}{2 m_{\beta \gamma}}\right)^2}}}},& \label{eq.A05}  \\
&\to  \frac{\Gamma_\beta}{2 m_{\beta \gamma} \ln q} \int_\Lambda^\infty {dk \frac{ \ln k} {k \left(\frac{q^{2}}{2 m_{\alpha \gamma}}+\frac{k^{2}}{2 m_{\beta \gamma}}\right)}}
 = \frac{\Gamma_\beta}{q^2 \ln q} \int_{\Lambda/q}^\infty {dy \frac{ \ln y+\ln q} {y \left(\frac{m_{\beta \gamma}}{m_{\alpha \gamma}}+y^2 \right)}} ,& \label{eq.A06}
\end{eqnarray}
where we changed integration variable, $k=q y$, in the last
expression.  Carrying out the two integrals we get
\begin{eqnarray}
&\int_{\Lambda/q}^\infty {dy \frac{ \ln y} {y \left(\frac{m_{\beta \gamma}}{m_{\alpha \gamma}}+y^2\right)}}=\left. \frac{1}{2} \frac{\ln^2 y}{(\frac{m_{\beta \gamma}}{m_{\alpha \gamma}}+y^2)}\right\vert_{\Lambda/q}^\infty &\nonumber\\
&+ \int_{\Lambda/q}^\infty {dy \frac{y \ln^2 y} { \left(\frac{m_{\beta \gamma}}{m_{\alpha \gamma}}+y^2\right)^2}}
\to -\frac{m_{\alpha \gamma}}{2m_{\beta \gamma}}\ln^2\left(\frac{\Lambda}{q}\right)
\to -\frac{\ln^2q}{2\frac{m_{\beta \gamma}}{m_{\alpha \gamma}}} ,& \label{eq.A07}
\end{eqnarray}
\begin{eqnarray}
&\int_{\Lambda/q}^\infty {dy \frac{1} {y \left(\frac{m_{\beta \gamma}}{m_{\alpha \gamma}}+y^2\right)}}= \left. \frac{\ln y}{(\frac{m_{\beta \gamma}}{m_{\alpha \gamma}}+y^2)}\right\vert_{\Lambda/q}^\infty  &\nonumber\\
&+ 2 \int_{\Lambda/q}^\infty {dy \frac{y \ln y} { \left(\frac{m_{\beta \gamma}}{m_{\alpha \gamma}}+y^2\right)^2}}\to -\frac{m_{\alpha \gamma}}{m_{\beta \gamma}}\ln \left(\frac{\Lambda}{q}\right)\to \frac{\ln q}{\frac{m_{\beta \gamma}}{m_{\alpha \gamma}}} ,& \label{eq.A08}
\end{eqnarray}
where we used that the integrals in the right-hand-side of
(\ref{eq.A07}) and (\ref{eq.A08}) are finite and their
contributions can be neglected when $q \to \infty$ in comparison with
the terms maintained.

In total, the spectator functions in (\ref{eq.A01}) are now found
by inserting (\ref{eq.A07}) and (\ref{eq.A08}) in (\ref{eq.A06}).  Notice that the 
contribution from (\ref{eq.A08}) has to be multiplied by  $\ln q$.  With the $\gamma-\beta$ interchange we also get the
second term in the right-hand-side of (\ref{eq.A01}).
The leading order large-momentum behavior of the spectator functions 
are therefore
\begin{equation}
\lim_{q \to \infty} f_{\alpha}\left(q\right) \to   \left(\frac{m_{\alpha \gamma}}{2m_{\beta \gamma}}\Gamma_\beta +\frac{m_{\alpha \beta}}{2m_{\beta \gamma}}\Gamma_\gamma\right) \frac{\ln q}{q^2} \ .
\label{eq.A09}
\end{equation} 
Replacing $f_\alpha(q_\alpha)$ in (\ref{eq.A09}) by its
conjectured asymptotic form, (\ref{eq.A00}), we find a system of
three linear equations for the three unknown, $\Gamma_\alpha =
\frac{m_{\alpha \gamma}}{2m_{\beta \gamma}}\Gamma_\beta
+\frac{m_{\alpha \beta}}{2m_{\beta \gamma}}\Gamma_\gamma$, which can
be rewritten as $ m_{\beta \gamma} \Gamma_\alpha = m_{\alpha \gamma}
\Gamma_\beta = \Gamma_\gamma m_{\alpha \beta}:=\Gamma$. The leading order
large-momentum asymptotic behavior for the three spectator functions
are then
\begin{equation}
\lim_{q \to \infty} f_\alpha(q) \to  \frac{\Gamma}{m_{\beta \gamma}}  \frac{\ln q}{q^2} \;.
\label{eq.A10}
\end{equation}
This result relates the asymptotic behavior of the three spectator
functions for one state.  The remaining constant $\Gamma$ still
depends on which excited state we consider, and furthermore also
on two-body masses and two-body energies.

\begin{figure}[!htb]%
\includegraphics[scale=0.5]{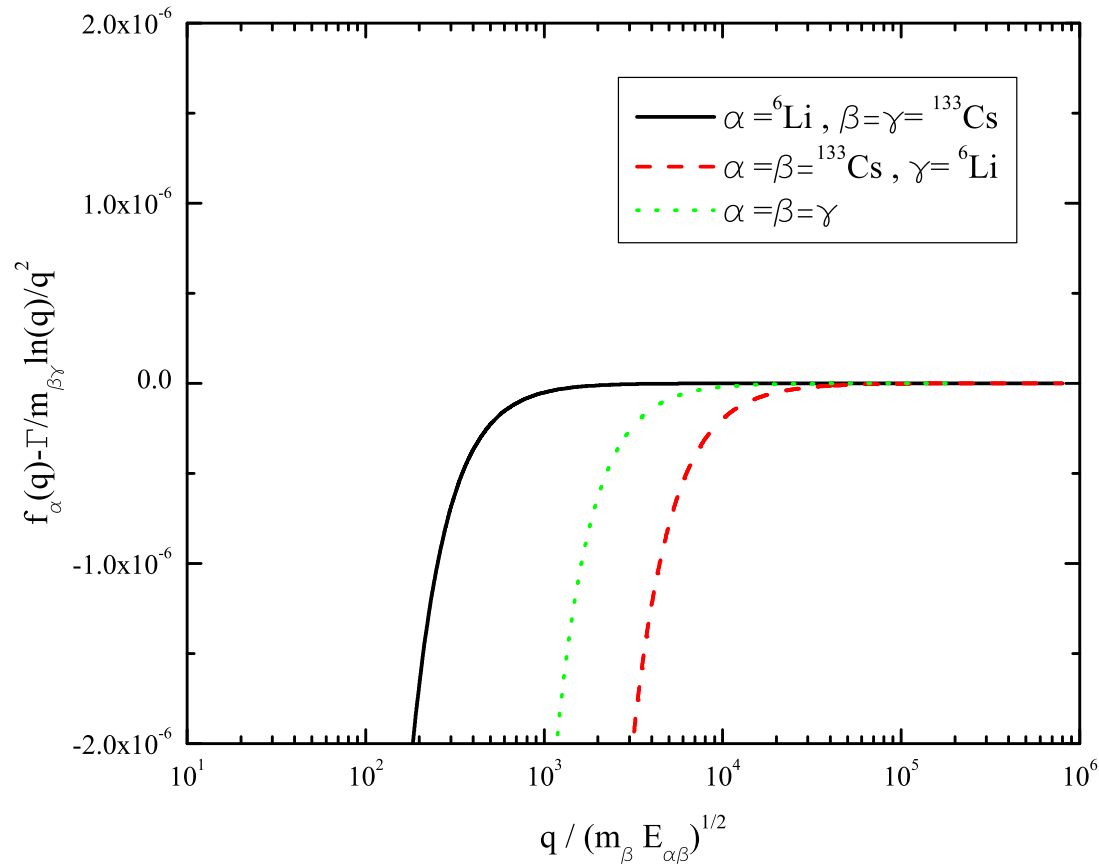}%
\caption{The difference $f_{\alpha}\left(q\right) -  \frac{\Gamma}{m_{\beta \gamma}} \frac{\ln q}{q^2}$ as a function of the momentum $q$. We see that (\ref{eq.A10}) exactly describes the asymptotic spectator function within our accuracy.}%
\label{fig.A01}%
\end{figure}

\begin{figure}[!htb]%
\includegraphics[scale=0.5]{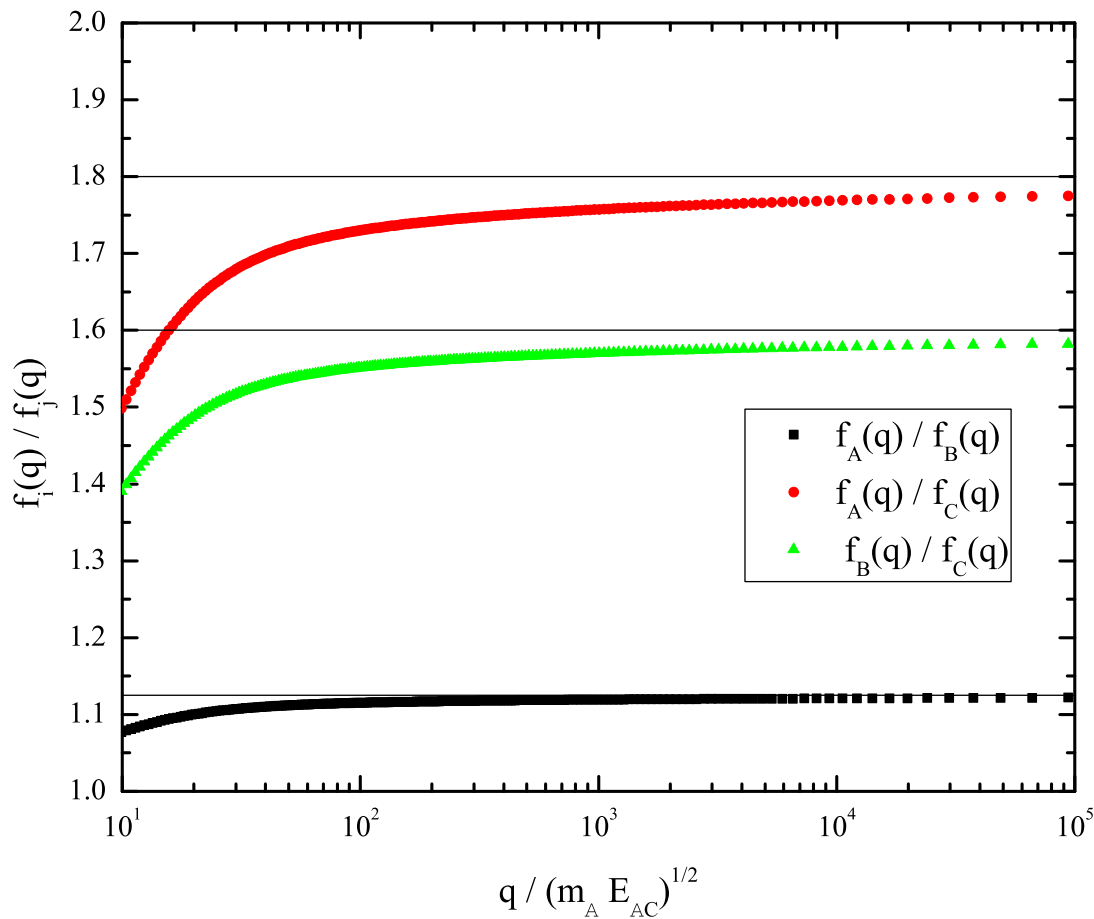}%
\caption{Ratios between the three distinct spectator function for a generic case of three distinct particles. Discrete points are the ratios between spectator functions numerically calculated through (\ref{eq.02}) and full lines are ratios between coefficients in (\ref{eq.A10}).}%
\label{fig.A02}%
\end{figure}

The derived large-momentum asymptotic behavior and the coefficients in
(\ref{eq.A10}) beautifully agree with the numerical calculation.
In figure \ref{fig.A01} we plot the difference
$f_{\alpha}\left(q\right) - \frac{\Gamma}{m_{\beta \gamma}} \frac{\ln
  q}{q^2}$ as a function of the momentum $q$ for the two different
spectator functions for the $^{133}$Cs-$^{133}$Cs-$^{6}$Li system. We
also show the same difference for a different system with three
identical bosons. First of all, this demonstrates that the large-momentum
behavior always is $\ln q/q^2$ for any $2D$ spectator function.
Secondly, for a given state the three cyclic permutations of
$f_{\alpha}\left(q\right) m_{\beta \gamma}$ for large $q$ approach the
same constant $\Gamma$ times $\ln q/q^2$.
This general large-momentum behavior is further demonstrated in figure
\ref{fig.A02} for a system of three distinct particles. The numerically
calculated points are compared to the full lines obtained from
(\ref{eq.A10}).  This comparison is again consistent with the
derived asymptotic behavior, and furthermore exhibit the rate and
accuracy of the convergence.  The limit is reached within 10\% and
1\% already for $\frac{q}{\sqrt{m_AE_{AC}}} \approx 50$ and 
$\frac{q}{\sqrt{m_AE_{AC}}} \approx 10^{4}$, respectively.

%
%

%
%

\subsection{Parametrizing from small to large momenta}
The asymptotic spectator function in (\ref{eq.A10}) seems to be a
good approximation even for moderate values of $q$, e.g. $q \approx 3
\sqrt{E_3}$.  We also have information about the large-distance
behavior for a given binding energy, that is $\exp(-\kappa \rho)$,
where $\kappa$ is related to the binding energy and $\rho$ is the
hyperradius.  Fourier transformation then relates to the small
momentum limit with an overall behavior of $(D+q^2)^{-1}$, where $D$
is a constant related to the energy.  This perfectly matches
(\ref{eq.01}) when two Jacobi momenta are present as in the
three-body system.  We therefore attempt a parametrization combining
the expected small momenta with the known large-momentum behavior:
\begin{equation}
f_\alpha(q)=f_\alpha(0)\frac{E_3}{\ln\sqrt{E_3}} \frac{\ln\left(\sqrt{\frac{q^2}{2 m_{\beta\gamma,\alpha}}+E_3}\right)}{\frac{q^2}{2 m_{\beta\gamma,\alpha}}+E_3} \ ,
\label{eq.A11}
\end{equation}   
where $f_\alpha(0)$ is a normalization constant which satisfies
$\int{d^2 q_\alpha\;n(q_\alpha)}=1$.

\begin{figure}[!htb]%
\includegraphics[scale=0.5]{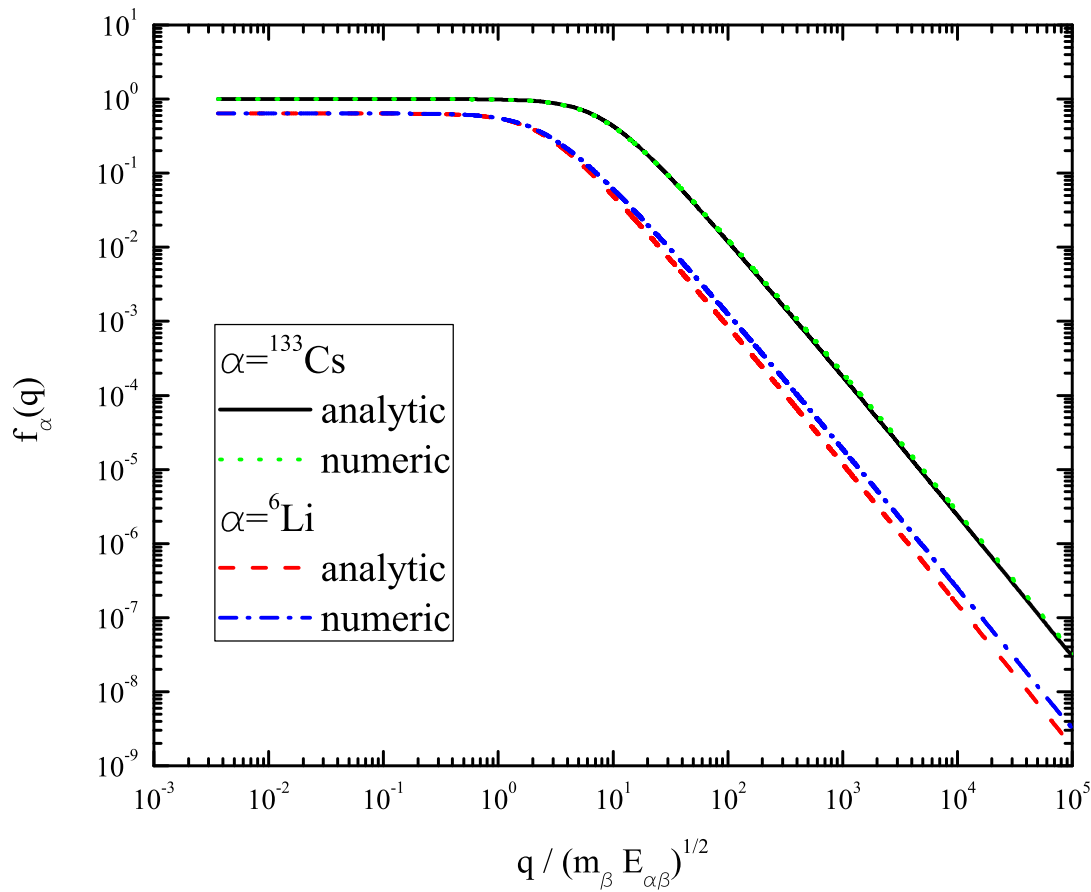}%
\caption{Comparison between the analytic spectator function estimated for the ground state, given in (\ref{eq.A11}) and the numeric solution of (\ref{eq.02}), for a $^{133}$Cs-$^{133}$Cs-$^{6}$Li system.  }%
\label{fig.A03}%
\end{figure}

We should first emphasize that excited states with the same angular
structure must have a different number of radial nodes. Therefore we
concentrate here only on the ground state.  The expression in
(\ref{eq.A11}) for the three ground state spectator functions
parametrizes the small momentum behavior almost perfectly for the case
of three distinguishable particles.  This is seen in Fig.\ref{fig.A03}
where we compare numerical and parametrized solution.  However, when
small momenta are reproduced the large-momentum limit deviates in
overall normalization, although with the same $q$-dependence.
Surprisingly, the analytic expression is most successful for the
spectator function related to the heaviest particle in the three-body
system.  This large-momentum mismatch is due to the normalization
choice in (\ref{eq.A11}), which is chosen to exactly reproduce the
$q=0$ limit.

%
%

\section{One-body large momentum density}\label{momentum}
The one-body density functions are observable quantities.  The most
directly measurable part is the limit of large momenta.  We therefore
separately consider the large-momentum limit of the four terms in
(\ref{eq.03a}) to (\ref{eq.03d}).  We employ the method sketched
in \cite{bel13a} and used to present numerical results for three
identical bosons in 2D.  Here we shall give more details and generalize
to systems of three distinguishable particles.

In three dimensions (3D), the corresponding problem was solved by inserting
the asymptotic spectator function (\ref{eq.A10}) into each of the
four terms in (\ref{eq.03a}) to (\ref{eq.03d}), and evaluating
the corresponding integrals \cite{castin2011,yam13}.  This procedure
is not guaranteed to work in $2D$  because momenta smaller than the asymptotic 
values may contribute in the integrands.  However, for $3D$ it was shown that
leading order in the integrands is sufficient to provide both leading
and next-to-leading order of the one-body momentum distributions.  
The
details of these calculation in $3D$ can be found in \cite{castin2011} for
three identical bosons and in \cite{yam13} for mass-imbalanced
systems.

The large-momentum spectator functions change a lot as dimensionality
is changed, going from $\sin(\ln(q))/q^2$ in 3D to $\ln(q)/q^2$ in
2D. If we try to naively proceed in 2D as successfully done in $3D$,
the integrals diverge.  We can circumvent this divergence problem by
following the procedure used in the derivation of the asymptotic
spectator functions.  In the following, we work out each of the four
momentum components defined in (\ref{eq.03a}) to (\ref{eq.03d}).
In addition we must simultaneously consider the next-to-leading order
term arising from the dominant $n_2$-term.

$\bullet n_1(q_\alpha)\;$: This term is straightforward to calculate.
The argument of the spectator function in (\ref{eq.03a}) does not
depend on the integration variable.  The large-momentum limit is then
found by replacing the spectator function by its asymptotic form and
taking the large $q$ limit after a simple integration.  We get then 
\begin{eqnarray}
\lim_{q_\alpha \to \infty} n_1(q_\alpha)& \to  4 \pi m_{\beta \gamma,\alpha} m_{\beta \gamma} \frac{\left|f_{\alpha}\left(q_\alpha\right)\right|^2} {q_\alpha^{2}} \to  4 \pi \frac{m_{\beta \gamma,\alpha}}{m_{\beta \gamma}} \Gamma^2 \frac{\ln^2 (q_\alpha)}{q_\alpha^6}. & 
\label{eq.04a}
\end{eqnarray}

$\bullet n_2(q_\alpha)\;$: We integrate the two terms in
(\ref{eq.03b}) over the angle as allowed by the simple structure
where the spectator function is angle independent.  The result 
\begin{eqnarray}
n_2(q_\alpha)= &2 \pi \int_0^\infty{dk \frac{k \left|f_{\beta}(k)\right|^2 \left( E_{3}+\frac{q_\alpha^{2}}{2m_{\alpha \gamma}}+\frac{k^{2}}{2m_{\beta \gamma}}\right)}{\left[\left( E_{3}+\frac{q_\alpha^{2}}{2m_{\alpha \gamma}}+\frac{k^{2}}{2m_{\beta \gamma}}\right)^2-\frac{k^2\;q_\alpha^2}{m_\gamma^2}\right]^{3/2}}}&\nonumber\\
& +
2 \pi \int_0^\infty{dk \frac{k \left|f_{\gamma}(k)\right|^2 \left( E_{3}+\frac{q_\alpha^{2}}{2m_{\alpha \beta}}+\frac{k^{2}}{2m_{\beta \gamma}}\right)}{\left[\left( E_{3}+\frac{q_\alpha^{2}}{2m_{\alpha \beta}}+\frac{k^{2}}{2m_{\beta \gamma}}\right)^2-\frac{k^2\;q_\alpha^2}{m_\beta^2}\right]^{3/2}}}&
\label{eq.B01}
\end{eqnarray}    
is then expanded for large $q$.  Since
$\int_0^\infty{dk\;k\;|f_\alpha(k)|^2}$ is finite, the large momentum
expansion becomes
\begin{eqnarray}
&\lim_{q_\alpha \to \infty} n_2(q_\alpha) \to  \frac{8 \pi}{q_\alpha^4} \left(m_{\alpha \gamma}^2 \int_0^\infty {dk\;k\;\left|f_{\beta}(k)\right|^2}+m_{\alpha \beta}^2 \int_0^\infty {dk\;k\;\left|f_{\gamma}(k)\right|^2} \right)&\nonumber\\
&+n_5(q_\alpha) 
\equiv  \frac{C_{\beta \gamma}}{q_\alpha^4}  + n_5(q_\alpha) \ , &
\label{eq.04b}
\end{eqnarray}
where the last equality defines $C_{\beta\gamma}$ which we call the 
two-body contact parameter for the $\beta\gamma$ two-body system.
The second term on the right-hand-side, $n_5(q_\alpha)$, gives the
next-to-leading term in the expansion of $n_2(q_\alpha)$.  It turns out
that this term has the same asymptotic behavior as $n_3(q_\alpha)$ and
$n_4(q_\alpha)$.   We must consequently keep it, but we postpone the 
derivation.
We emphasize that the one-body large-momentum leading order comes only
from $n_2(q_\alpha)$.  Here we cannot replace the spectator function
by its asymptotic expression, because the main contribution to
$\int_0^\infty{dk\;k\;|f_\alpha(k)|^2}$ arises from small $k$.  This
replacement would therefore lead to a completely wrong result.
However, this is not always the case, as we shall see later for
$n_5(q_\alpha)$.

$\bullet n_3(q_\alpha)\;$: The structure of $n_3(q_\alpha)$ in
(\ref{eq.03c}) is similar to $n_2(q_\alpha)$ in
(\ref{eq.03b}). The only difference is that the spectator function
under the integration sign now is not squared. This small functional
difference leads to a completely different result. As in the previous
case, we can still carry out the angular integration, which only
involves the denominator. Integrating (\ref{eq.03c}) over the
angle we get
\begin{eqnarray}
n_3(q_\alpha)= 4 \pi f_{\alpha}(q_\alpha)&\left(\int_0^\infty {dk \frac{k f_{\beta}(k) \left( E_{3}+\frac{q_\alpha^{2}}{2m_{\alpha \gamma}}+\frac{k^{2}}{2m_{\beta \gamma}}\right)}{\left[\left( E_{3}+\frac{q_\alpha^{2}}{2m_{\alpha \gamma}}+\frac{k^{2}}{2m_{\beta \gamma}}\right)^2-\frac{k^2\;q_\alpha^2}{m_\gamma^2}\right]^{3/2}}}\right. &\nonumber\\
&\left.+\int_0^\infty{dk \frac{k f_{\gamma}(k) \left( E_{3}+\frac{q_\alpha^{2}}{2m_{\alpha \beta}}+\frac{k^{2}}{2m_{\beta \gamma}}\right)}{\left[\left( E_{3}+\frac{q_\alpha^{2}}{2m_{\alpha \beta}}+\frac{k^{2}}{2m_{\beta \gamma}}\right)^2-\frac{k^2\;q_\alpha^2}{m_\beta^2}\right]^{3/2}}} \right)\ . &
\label{eq.B02}
\end{eqnarray}    
Here, the difference between $n_2$ and $n_3$ becomes important, since
$\int_0^\infty{dk\;k\;f(k)}$ is divergent and we can not expand
(\ref{eq.B02}) as we did for (\ref{eq.B01}).  We shall instead
proceed as done in obtaining the asymptotic spectator function.  We
divide the integration in (\ref{eq.B02}) at a large, but finite,
momentum, $\Lambda \gg \sqrt{E_3}$, and each term on the
right-hand-side is split in two others.  The two terms only differ by
simple factors, and we therefore only give details for the first term.
Changing variables to $k=q_\alpha y$, (\ref{eq.B02}) becomes
\begin{eqnarray}
&\lim_{q_\alpha \to \infty} n_3(q_\alpha) \to&\nonumber\\  
&16 \pi m_{\beta\gamma}^2 \frac{f_{\alpha}(q_\alpha)}{q_\alpha^2} \int_0^{\Lambda/q_\alpha}{dy \frac{y f_{\beta}(q_\alpha y) \left( \frac{2 m_{\beta \gamma} E_{3}}{q_\alpha^{2}}+\frac{m_{\beta \gamma}}{m_{\alpha \gamma}}+y^2 \right)}{\left[\left( \frac{2 m_{\beta \gamma} E_{3}}{q_\alpha^{2}}+\frac{m_{\beta \gamma}}{m_{\alpha \gamma}}+y^2 \right)^2-\frac{4 m_{\beta \gamma}^2}{m_\gamma^2}y^2\right]^{3/2}}} & \nonumber\\ 
&+ 16 \pi m_{\beta\gamma}^2 \frac{f_{\alpha}(q_\alpha)}{q_\alpha^4} \frac{\Gamma}{m_{\alpha\gamma}} \int_{\Lambda/q_\alpha}^\infty{dy \frac{ [\ln(q_\alpha)+\ln(y)] \left( \frac{m_{\beta \gamma}}{m_{\alpha \gamma}}+y^2 \right)}{y\;\left[\left(\frac{m_{\beta \gamma}}{m_{\alpha \gamma}}+y^2 \right)^2-\frac{4 m_{\beta \gamma}^2}{m_\gamma^2}y^2\right]^{3/2}}} + ...\ ,&
\label{eq.B03}
\end{eqnarray}
where $f_\beta(k)$ is replaced by its asymptotic form and $E_3$ is
neglected in the second term on the right-hand-side, where $\sqrt{E_3}
\ll \Lambda$ and $k>\Lambda$.  In the limit $q_\alpha \to \infty$, the
integral vanishes in the first term which therefore does not
contribute to the large-momentum limit. The integrals in the second
term are
\begin{eqnarray}
&\int_{\Lambda/q_\alpha}^\infty{dy \frac{ \ln(y)\; h(y)}{y}}=\left.\frac{1}{2} \ln^2(y)\;h(y) \right\vert_{\Lambda/q_\alpha}^\infty -\frac{1}{2} \int_{\Lambda/q_\alpha}^\infty {dy  \ln^2(y) g(y)} \to &\nonumber\\
&-\frac{m_{\alpha \gamma}^2}{2m_{\beta \gamma}^2}\ln^2\left(\frac{\Lambda}{q_\alpha}\right)\to -\frac{\ln^2(q_\alpha)}{2\frac{m_{\beta \gamma}^2}{m_{\alpha \gamma}^2}} ,& \label{eq.B04}
\end{eqnarray}
\begin{eqnarray}
&\int_{\Lambda/q_\alpha}^\infty{dy \frac{h(y)}{y}}=\left. \ln(y) h(y) \right\vert_{\Lambda/q_\alpha}^\infty -\int_{\Lambda/q_\alpha}^\infty {dy  \ln(y) g(y)} \to &\nonumber\\
&-\frac{m_{\alpha \gamma}^2}{m_{\beta \gamma}^2}\ln\left(\frac{\Lambda}{q_\alpha}\right)\to \frac{\ln(q_\alpha)}{\frac{m_{\beta \gamma}^2}{m_{\alpha \gamma}^2}} ,& \label{eq.B05}
\end{eqnarray}
where 
\begin{eqnarray}
h(y)= \left(\frac{m_{\beta \gamma}}{m_{\alpha
    \gamma}}+y^2\right)\left[\left(\frac{m_{\beta \gamma}}{m_{\alpha
      \gamma}}+y^2 \right)^2-\frac{4 m_{\beta
      \gamma}^2}{m_\gamma^2}y^2\right]^{-3/2} , \\
g(y)=\frac{d\;h(y)}{dy},\;\; \lim_{y \to 0} \ln^2(y) g(y) \to 0\;,\;\;
\lim_{y \to \infty} \ln^2(y) g(y) \to 0 \;.
\end{eqnarray}
The function $g(y)$ and its limits ensure that the integrals on the
right-hand-side of Eqs. (\ref{eq.B04}) and (\ref{eq.B05}) are finite
and their contributions to the momentum distribution can be neglected
when $q_\alpha \to \infty$.
Finally, inserting the results given in (\ref{eq.B04}) and
(\ref{eq.B05}) into (\ref{eq.B03}) and replacing the spectator
function $f_\alpha(q_\alpha)$ by its asymptotic form, the
leading order term of the one-body momentum 
distribution from $n_3(q_\alpha)$ is given by
\begin{equation}
\lim_{q_\alpha \to \infty} n_3(q_\alpha) \to  8 \pi \left(\frac{m_{\alpha\gamma}+m_{\alpha \beta}}{m_{\beta\gamma}}\right) \Gamma^2 \frac{\ln^3(q_\alpha)}{q_\alpha^6} \ ,
\label{eq.04c}
\end{equation}
where the second term in the right-hand-side of (\ref{eq.B02}) is
recovered and added by the interchange of 
$m_{\alpha \gamma} \to m_{\alpha \beta}$ in (\ref{eq.B03}) to (\ref{eq.B05}).

Although $n_2(q_\alpha)$ and $n_3(q_\alpha)$ have rather similar form,
their contributions to the one-body large momentum density are quite
different. As we shall see later, the next-to-leading order,
$n_5(q_\alpha)$, of $n_2(q_\alpha)$ is comparable to the
$n_3(q_\alpha)$ leading order, given in (\ref{eq.04c}).

$\bullet n_4(q_\alpha)\;$: This is the most complicated of the four
additive terms in the one-body momentum density. The angular
dependence in both spectator arguments cannot be removed
simultaneously by variable change.  The formulation in
(\ref{eq.03d}) has the advantage that the argument in
$f_\gamma(|\mathbf{k+q_\alpha}|)$ (or in
$f_\beta(|\mathbf{k+q_\alpha}|)$) is never small in the limit of large
$q_\alpha$.  This is in contrast to a choice of variables where the
numerator in the first term of (\ref{eq.03d}) would be
$f_{\gamma}(k)f_{\beta}(|\mathbf{k-q_\alpha}|)$, and the argument in
$f_\beta$ would consequently be small as soon as $k$ is comparable to
$q_\alpha$.
The main contribution to the integrals in (\ref{eq.03d}) arise
from small $k$.  For large $q_\alpha$, we can then use the
approximation, $f_\gamma(|\mathbf{k+q_\alpha}|) \approx
f_\gamma(q_\alpha)$ (or $f_\beta(|\mathbf{k+q_\alpha}|) \approx
f_\beta(q_\alpha)$).  The integrals are then identical to the
terms of $n_3$ in (\ref{eq.03c}). By keeping track of the slightly
different mass factors we therefore immediately get the asymptotic
limit to be
\begin{equation}
\lim_{q_\alpha \to \infty} n_4(q_\alpha) \to  4 \pi \Big(\frac{m_{\alpha\gamma}}{m_{\alpha \beta}} 
+ \frac{m_{\alpha\beta}}{m_{\alpha \gamma}} \Big)
\Gamma^2 \frac{\ln^3(q_\alpha)}{q_\alpha^6} \ .
\label{eq.04d}
\end{equation}

$\bullet n_5(q_\alpha)\;$: This is the next-to-leading order
contribution from the $n_2(q_\alpha)$ term.  It turns out that this
term has the same large-momentum behavior as the leading orders of
both $n_3(q_\alpha)$ and $n_4(q_\alpha)$. By definition we have
\begin{equation} \label{eq.n5a}
n_5(q_\alpha) =  n_2(q_\alpha) - \lim_{q_\alpha \to \infty} n_2(q_\alpha)
 = n_2(q_\alpha) -\frac{C_{\beta \gamma}}{q_\alpha^4}
\end{equation}
which can be rewritten in details as
\begin{eqnarray}
n_5(q_\alpha) = &\lim_{q_\alpha \to \infty}
 2 \pi \int_0^\infty{dk k \left|f_{\beta}(k)\right|^2} &\nonumber\\
& \times\left(\frac{
\left( E_{3}+\frac{q_\alpha^{2}}{2m_{\alpha \gamma}}+\frac{k^{2}}{2m_{\beta \gamma}}\right)}{\left[\left( E_{3}+\frac{q_\alpha^{2}}{2m_{\alpha \gamma}}+\frac{k^{2}}{2m_{\beta \gamma}}\right)^2-\frac{k^2q_\alpha^2}{m_\gamma^2}\right]^{3/2}} - \frac{4 m_{\alpha \gamma}^2}{q_\alpha^4}\right) + ....& \;,
\label{eq.n5b}
\end{eqnarray}    
where the dots denote the last term in (\ref{eq.B01}) obtained by
interchange of $\beta$ and $\gamma$.  The tempting procedure is now to
expand the integrand around $q_{\alpha} = \infty$ assuming that
$q_{\alpha}$ overwhelms all terms in this expression.  This
immediately leads to integrals corresponding to the cubic moment of
the spectator function which however is not converging.  On the other
hand (\ref{eq.n5b}) is perfectly well defined due to the large-$k$
cut-off from the denominator.  In fact, the spectator function is
multiplied by $k^3$ and $1/k^{3}$ at small and large $k$-values,
respectively.  The integrand therefore has a maximum where the main
contribution to $n_5$ arises. This peak in $k$ moves towards infinity
proportional to $q$.
To compute $n_5(q_\alpha)$ we then divide the integration into two
intervals, that is from zero to a finite but very large $k$-value,
$\Lambda_s$, and from $\Lambda_s$ to infinity.  The small momentum
interval, $k/q_{\alpha} \ll 1$, allows an expansion in $k/q_{\alpha}$
leading to the following contribution $n_{5,1}(q_{\alpha})$:
\begin{equation}
n_{5,1}(q_\alpha) = 8 \pi \frac{m_{\alpha \gamma}^2}{q_\alpha^6} 
 \left( 3 \frac{m^{2}_{\alpha \gamma}}{m_\gamma^2}
- \frac{m_{\alpha \gamma}}{m_{\beta \gamma}}\right)
\int_0^{\Lambda_s}dk k^3 \left|f_{\beta}(k)\right|^2 
+ \frac{\eta}{q_{\alpha}^8}+\dots \;,
\label{eq.n5c}
\end{equation}    
where $\eta$ is a constant. 
Thus the contribution from this small momentum integration vanish with
the $6^{\rm th}$ power of $q_{\alpha}$ which is faster than the other
sub-leading orders we want to keep.
We choose $\Lambda_s$ sufficiently large for the spectator function to
reach its asymptotic behavior in (\ref{eq.A10}).  The large
interval integration can now be performed by omitting the small
$E_3$-terms and change of integration variable to $y$, $k^2=y q_{\alpha}^2$, i.e. 
\begin{eqnarray}
n_{5,2}(q_\alpha) = &\frac{ \pi \Gamma^2}{q_{\alpha}^6}
\int_{\Lambda_s^2/q_{\alpha}^2}^{\infty} \frac{dy}{y^2} 
\left[\ln^2(y) + \ln^2(q_{\alpha}^2) + 2 \ln y \ln (q_{\alpha}^2)\right]&\nonumber\\
&\times\left(\frac{1 + y m_{\alpha \gamma}/ m_{\beta \gamma}}
{\left[(1 + y m_{\alpha \gamma}/ m_{\beta \gamma})^2 -
 4 y m^2_{\alpha \gamma}/ m^2_{\gamma}\right]^{3/2}} -1\right) + .... &\;,
\label{eq.n5d}
\end{eqnarray}
where the large $y$-limit behaves like $\ln^2 y/y^4$ and therefore
assuring rapid convergence, whereas the integrand for small $y$
behaves like $(\ln^2(y) + \ln^2(q^2) + 2 \ln y \ln (q^2))/y$. Thus,
the integration from an arbitrary minimum value, $y_L$ (independent of $q_\alpha$), of $y>
\Lambda_s^2/q_{\alpha}^2$ yields a $q_{\alpha}$-independent value
except for the logarithmic factors and $q_{\alpha}$ in the
numerator. Thus the large-$q_{\alpha}$ dependence is found from very
small values of $y$ close to the lower, and vanishing, limit.  In
total we get by expansion in small $y$ that the limit for large
$q_{\alpha}$ approaches zero as
\begin{eqnarray}
\lim_{q_\alpha \to \infty} n_{5,2}(q_\alpha) \to & \frac{16 \pi \Gamma^2}{q_{\alpha}^6}
 \left( 3 \frac{m^{2}_{\alpha \gamma}}{m_\gamma^2}
- \frac{m_{\alpha \gamma}}{m_{\beta \gamma}}\right)&\nonumber\\
&\int_{\Lambda_s^2/q_\alpha^2}^{y_L} \frac{dy}{y} 
[\ln^2(y) + \ln^2(q_{\alpha}^2) + 2 \ln y \ln (q_{\alpha}^2)] \; .&
\label{eq.n5e}
\end{eqnarray}
Together with the term from interchange of $\beta$ and $\gamma$ in
(\ref{eq.A10}) we get in total that
\begin{equation}
\lim_{q_\alpha \to \infty} n_{5}(q_\alpha) \to  \frac{16 \pi}{3}
 \left[ 3 \left( \frac{m^{2}_{\alpha \gamma}}{m_\gamma^2}+\frac{m^{2}_{\alpha \beta}}{m_\beta^2} \right)
- \frac{m_{\alpha \gamma} + m_{\alpha \beta}}{m_{\beta \gamma}}\right]  \Gamma^2 \frac{\ln^3(q_\alpha)}{q_\alpha^6}\; .
\label{eq.n5f}
\end{equation}

\section{Two- and Three-body contact parameters}\label{contact}
We first collect the analytically derived relations, and secondly we
compare to numerically calculated values.

\subsection{Analytic expressions}
Two- and three-body contact parameters are defined via the
large-momentum one-body density. The two-body contact parameter,
$C_{\beta\gamma}$, is the proportionality constant of the leading
order $q_\alpha^{-4}$ term, which arises solely from $n_2(q_\alpha)$
in (\ref{eq.04b}).  For three distinguishable particles we have
three contact parameters each related to the momentum distribution of
one particle.  The definition is already given in (\ref{eq.04b}).
They are related through
\begin{equation}
C_{\alpha \beta} + C_{\alpha \gamma} = C_{\beta \gamma} + 16 \pi m_{\beta \gamma}^2 \int_0^\infty {dk\;k\;\left|f_{\alpha}(k)\right|^2} \ ,
\label{eq.06}
\end{equation}
and cyclic permutations.
For a specific system, where two of the particles are non-interacting
in 2D, the corresponding two-body energy vanishes, $E_{\beta
  \gamma}=0$ \cite{bel11,bel12}. Then, from (\ref{eq.02}) the
spectator function also vanishes, $f_\alpha(q)=0$, and
(\ref{eq.06}) reduces to
\begin{equation}
C_{\alpha \beta} + C_{\alpha \gamma} = C_{\beta \gamma} \;\;\textrm{for}\;\; E_{\beta\gamma}=0 \ .
\label{eq.08}
\end{equation}
In this case, we have this simple relation between the three two-body
contact parameters.  This relation between different two-body
parameters does not depend on system dimension.  Although our
calculations are in 2D, this relation in (\ref{eq.08}) applies as
well for 3D systems with one non-interacting subsystem.  We emphasize
that a non-interacting system and a vanishing two-body energy is not
the same in $3D$ where some attraction is necessary to provide a state
with zero binding energy.

The three-body contact parameter, $C_{\beta \gamma,\alpha}$, is
defined as the proportionality constant on the next-to-leading order
in the one-body large-momentum density distribution.  For
distinguishable particles we have again three of these parameters,
each related to one of the particle's momentum distributions.  The
asymptotic behavior, $\ln^3(q_\alpha)/q_\alpha^6$, receives
contributions from the three terms specified in
(\ref{eq.04c}), (\ref{eq.04d}) and (\ref{eq.n5f}).  In total
we have
\begin{equation}
C_{\beta \gamma,\alpha} = 16 \pi \left( \frac{m_{\alpha \gamma}+m_{\alpha \beta}}{6 m_{\beta \gamma}}+\frac{m_{\alpha \gamma}}{4m_{\alpha \beta}} +
\frac{m_{\alpha \beta}}{4m_{\alpha \gamma}}
 + \frac{m_{\alpha \gamma}^2}{m_\gamma^2}+\frac{m_{\alpha \beta}^2}{m_\beta^2}\right) \Gamma^2 \ .
\label{eq.07}
\end{equation}
It is worth emphasizing that only a logarithmic factor distinguishes
the behavior of the three-body contact term from the next order,
$\ln^2(q_\alpha)/q_\alpha^6$ which arises from $n_1$ as well as from
$n_2$, $n_3$, and $n_5$.  In practical measurements, it must be a
huge challenge to distinguish between terms differing by only one
power of $\ln(q_\alpha)$.

For the three-body contact parameter, (\ref{eq.07}) with only one
non-interacting two-body system, we get
\begin{equation}
C_{\beta \gamma,\alpha} = 16 \pi \left( - \frac{m_{\alpha \gamma}+m_{\alpha \beta}}{3 m_{\beta \gamma}}+\frac{m_{\alpha \gamma}}{4m_{\alpha \beta}} +
\frac{m_{\alpha \beta}}{4m_{\alpha \gamma}}
+ \frac{m_{\alpha \gamma}^2}{m_\gamma^2}+\frac{m_{\alpha \beta}^2}{m_\beta^2}\right) \Gamma^2 \ ,
\label{eq.07a}
\end{equation}
which is obtained by collecting contributions from only the 
non-vanishing $n_4$ and $n_5$ terms (since $f_\alpha(q)=0$, $n_1$ and
$n_3$ do not contribute).
Cyclic permutations of the indices in (\ref{eq.07}) and
(\ref{eq.07a}) now allow the conclusion that the three different
three-body contact parameters are related by the mass factors in
(\ref{eq.07}) and (\ref{eq.07a}).  This conclusion holds for all
excited states. Universality of independence of excited state is
another matter and in fact not found numerically.

\subsection{Numerical results}
The results in the preceding subsection hold for any mass-imbalanced
three-body system. Such a system has six independent parameters, which
are reduced to four by choosing one mass and one energy as units
\cite{bel12}.  This merely implies that all results can be expressed
as ratios of masses and energies, and in this way providing very
useful scaling relations.  Results depending on four independent
parameters are still hard to display and digest.

To built up our understanding, we now focus on systems composed of two
identical particles, $A$, and a distinct one, $C$.  Such a system has
four independent parameters from the beginning, which are reduced to two
after choice of units.  From now on we shall use $E_{AC}$ and $m_A$ as
our energy and mass units, and for simplicity we introduce the mass
ratio $m=\frac{m_C}{m_A}$.  In these units energies
and momenta appearing in the equations must be multiplied by
$E_{AC}$ and $\sqrt{m_A E_{AC}}$, respectively.  For this system the two-body contact
  parameters in (\ref{eq.06}) are given by
\begin{eqnarray}
&C_{AA}=16 \pi \left(\frac{m}{1+m}\right)^2 \int_0^\infty {dk\;k\;\left|f_{A}(k)\right|^2} \ ,& \label{eq.09a} \\
&C_{AC}= \frac{C_{AA}}{2}+ 2 \pi \int_0^\infty {dk\;k\;\left|f_{C}(k)\right|^2} \ ,& \label{eq.09b}
\end{eqnarray}

For three identical particles where all masses and interactions are
the same, $C_{AA}=C_{AC} = C_2$, and the quantity $\frac{C_2}{E_3}$ is
a universal constant in 2D  \cite{bel13a,werner2012b}.  To be
explicit, this quantity has the same value for the only two existing
bound states, ground and first excited state.  Mass-imbalanced systems
have a richer energy spectrum with many excited states
\cite{bel12,bel13b}.  Maintaining the universal conditions for all
excited states is obviously more demanding.  

\begin{figure}[!htb]%
\includegraphics[scale=0.5]{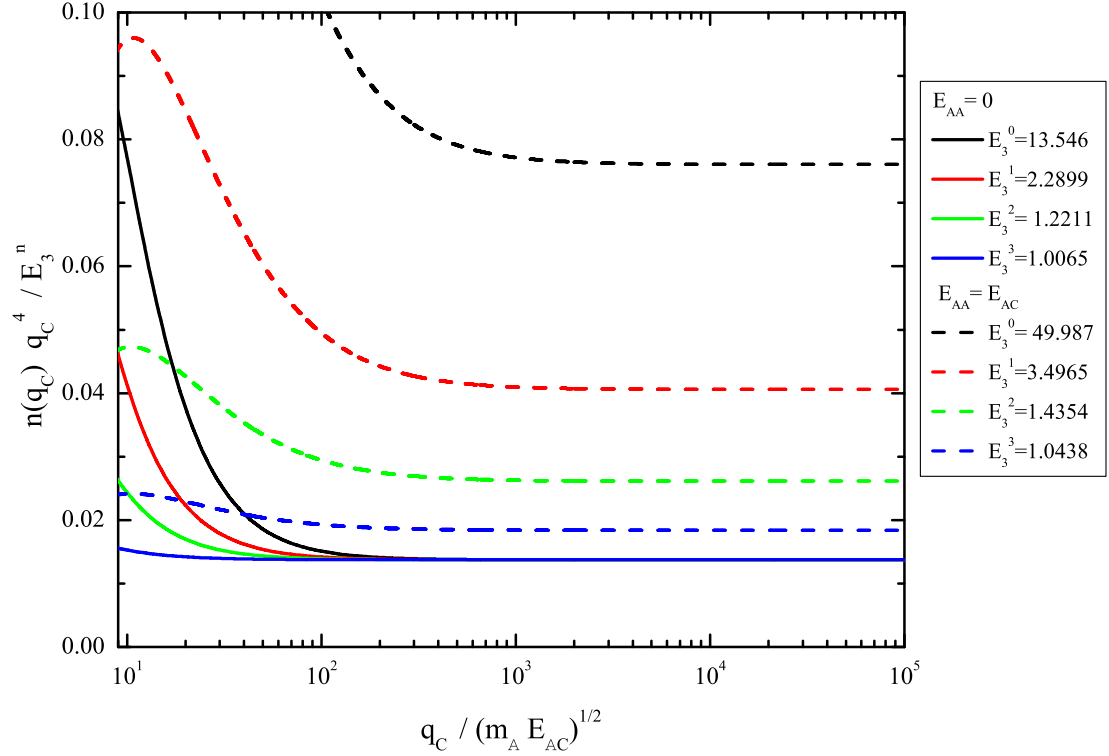}%
\caption{
The leading order of the one-body momentum density divided by $E_3^n$ for each bound state labeled as $n$ in a system composed of two identical ( $A=^{133}$Cs ) particles and a distinct one ( $C=^{6}$Li ) as a function of the momentum $q$ for both $E_{AA}=1$ and $E_{AA}=0$.}
\label{fig.07}%
\end{figure}

Detailed investigations reveal that when the mass-energy symmetry is
broken, the universality of $\frac{C_2}{E_3}$ does not hold any more.
The two two-body contact parameters defined in (\ref{eq.09a}) and
(\ref{eq.09b}) divided by the three-body energy are not the same for
all possible bound states in the general case.  However, in at least
one special case of two identical non-interacting particles,
$E_{AA}=0$, the universality is recovered.  This is implied by
$f_C=0$ as seen from the set of coupled homogeneous integrals
equations (\ref{eq.02}).  Then the two universal two-body contact
parameters are related, that is
\begin{eqnarray}
&C_{AC}= \frac{C_{AA}}{2} \;\;\textrm{for}\;\; E_{AA}=0 \ .& \label{eq.10}
\end{eqnarray}
We illustrate in Fig.~\ref{fig.07} how the two-body contact parameters
vary with excitation energy for a mass-asymmetric system.  We choose
$^{133}$Cs$-^{133}$Cs$-^{6}$Li corresponding to $A=^{133}$Cs and
$C=^{6}$Li.  This system has four excited states at energies depending
on the size of $E_{AA}$, and the large-momentum limit of constants is
reached in all cases.  For $E_{AA}=0$, universality is observed, since
all two-body contact values, $C_{AC}/E_3$ , are equal in units of the
three-body energy.  This case is rather special because two particles
do not interact and the three-body structure is determined by the
identical two-body interactions between the other two subsystems.  In
other words the large-momentum limit of particle $A$ is determined
universally by the properties of the $A-C$ subsystem.  The
other contact parameter, $C_{AA}/E_3$, is also universal and following
from (\ref{eq.10}).

This picture changes when all particles interact as seen in
Fig.~\ref{fig.07} for $E_{AA}=1$. Now the large-momentum limit, still
constants independent of momentum, changes with the excitation energy.
The systematics is that both $C_{AA}/E_3$ and $C_{AC}/E_3$ as function
of excitation energy move towards the corresponding values for
$E_{AA}=0$, one from below and the other from above.  First the
non-universality is understandable, since the interaction of the two
identical particles now must affect the three-body structure at small
distances, and hence at large momenta.  However, as the three-body
binding energy decreases, the size of the system increases and details
of the short-distance structure becomes less important. 

The quantities $\frac{C_{AA}}{E_3}$ and $\frac{2 \pi}{E_3}
\int_0^\infty {dk\;k\;\left|f_{C}(k)\right|^2}$ are defined by the
limiting large-$q$ behavior of $n_2$ in (\ref{eq.04b}).  Plotting
the corresponding pieces of $n_2(q) q^4$ as function of $q$ lead to
figures much similar to Fig.~\ref{fig.07}, where different excitation
dependent lines emerge for $E_{AA}=1$, while they all coincide for
$E_{AA}=0$.  The constant values for $\frac{C_{AC}}{E_3}$,
$\frac{C_{AA}}{E_3}$ and $\frac{2 \pi}{E_3} \int_0^\infty
{dk\;k\;\left|f_{C}(k)\right|^2}$ in the limit $q \to \infty$, are
shown in Table~\ref{tab1}, for two different interactions and two
different systems represented by $C=^{6}$Li and $A=^{133}$Cs or
$A=^{39}$K.  These results of numerical calculations confirm the
systematics described above in complete agreement with
(\ref{eq.09b}) and (\ref{eq.10}).

\begin{table}
\centering
\caption{The constant values for the scaled two-body contact
  parameters $\frac{C_{AC}}{E_3}$,
  $\frac{C_{AA}}{E_3}$ and $\frac{2 \pi}{E_3} \int_0^\infty
  {dk\;k\;\left|f_{C}(k)\right|^2}$ in the limit $q \to \infty$, are
  shown in table~\ref{tab1}, for two different interactions and two
  different systems represented by $C=^{6}$Li and $A=^{133}$Cs or
  $A=^{39}$K.  The values in the fifth column are plotted in
  Fig.~\ref{fig.07}.}
\begin{tabular}{|c|c|c|c|c|c|}
\hline
system & $\frac{E_{AA}}{E_{AC}}$ & state  & $\frac{C_{AA}}{E_3}$ & $\frac{C_{AC}}{E_3}$ & $\frac{2 \pi}{E_3} \int_0^\infty {dk\;k\;\left|f_{C}(k)\right|^2}$ \\ \hline
\multirow{5}{*}{$A=^{133}$Cs $C=^{6}$Li}  
	& \multirow{4}{*}{1} 
		& Ground & 0.02210  & 0.07625  & 0.06503 \\ \cline{3-6}
		&& First  & 0.02495  & 0.04062  & 0.02812 \\ \cline{3-6}
		&& Second & 0.02616  & 0.02612  & 0.01305 \\ \cline{3-6}
		&& Third  & 0.02718  & 0.01837  & 0.00478 \\ \cline{2-6}
	&\multirow{1}{*}{0}
		& all & 0.02748  & 0.01374  & 0 \\ \hline 

\multirow{4}{*}{$A=^{39}$K  $C=^{6}$Li}
	&\multirow{3}{*}{1}
		& Ground & 0.06337  & 0.11499  & 0.08372 \\ \cline{3-6}
		&& First  & 0.07438  & 0.08256  & 0.04727 \\ \cline{3-6}
		&& Second & 0.07934  & 0.05369  & 0.01840 \\ \cline{2-6}
	&\multirow{1}{*}{0}
		& all & 0.08304  & 0.04152  & 0 \\ \hline
\end{tabular}
\label{tab1}
\end{table}

\begin{figure}[!htb]%
\includegraphics[scale=0.5]{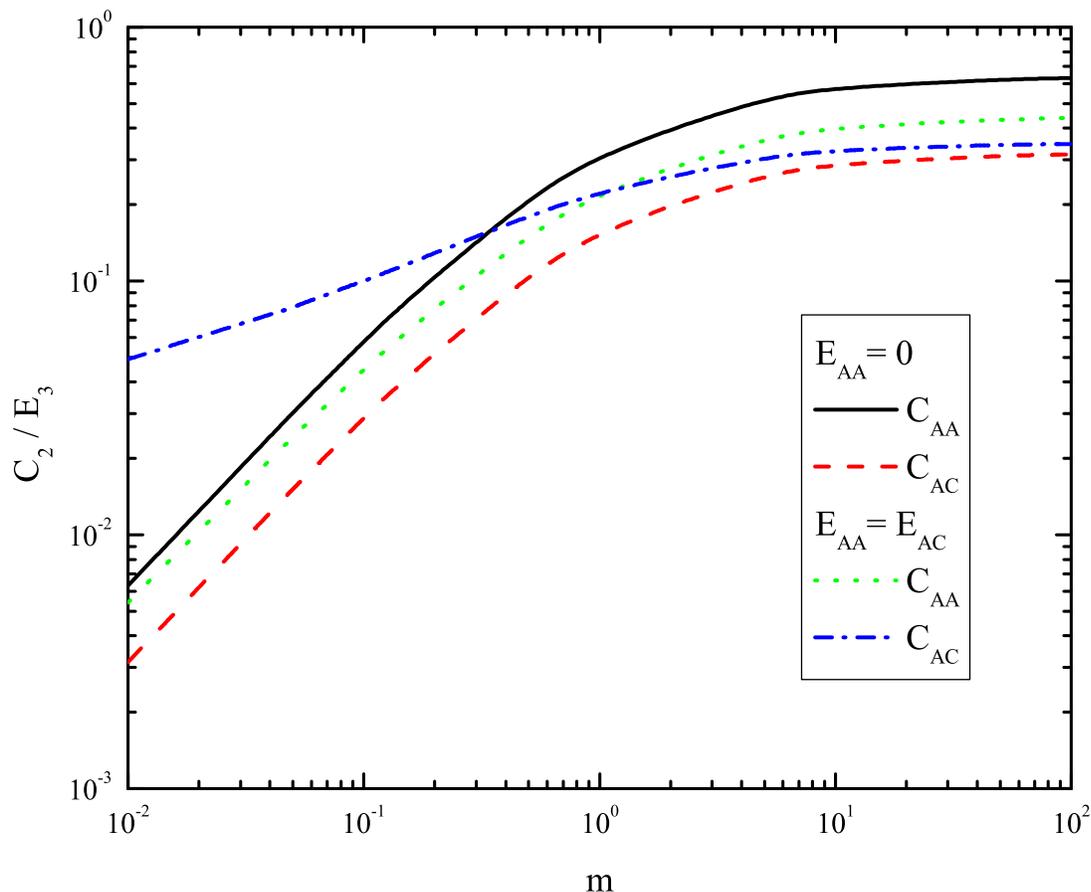}%
\caption{The two-body parameters $C_{AA}$ and $C_{AC}$ defined in (\ref{eq.04b}) as function of the mass ratio $m=\frac{m_C}{m_A}$ for an $AAC$ system in both cases where $E_{AA}=0$ and $E_{AA}=E_{AC}$.}%
\label{fig.10}%
\end{figure}

In general, for two identical particles the two-body contact
parameters divided by the three-body energy depend on the mass ratio
$m$.  The dependence change from universal for $E_{AA}=0$ to
non-universal for $E_{AA}=1$.  The mass dependence for ground states
is shown in Fig.~\ref{fig.10}, where we see that the ratio
$\frac{C_{AA}}{C_{AC}}=2$ in (\ref{eq.10}) holds for $E_{AA}=0$ in
the entire mass interval investigated.  We also see how the second term
on the right-hand-side of (\ref{eq.09b}) affects the relation
between the two two-body contact parameters.  Fig.~\ref{fig.10} shows
that the values rapidly increase from small $m$ up to
$1$ and become almost constant above $m \approx 5$.  This behavior is
similar to mass-imbalanced system in 3D \cite{yam13}.

\begin{figure}[!htb]%
\includegraphics[scale=0.5]{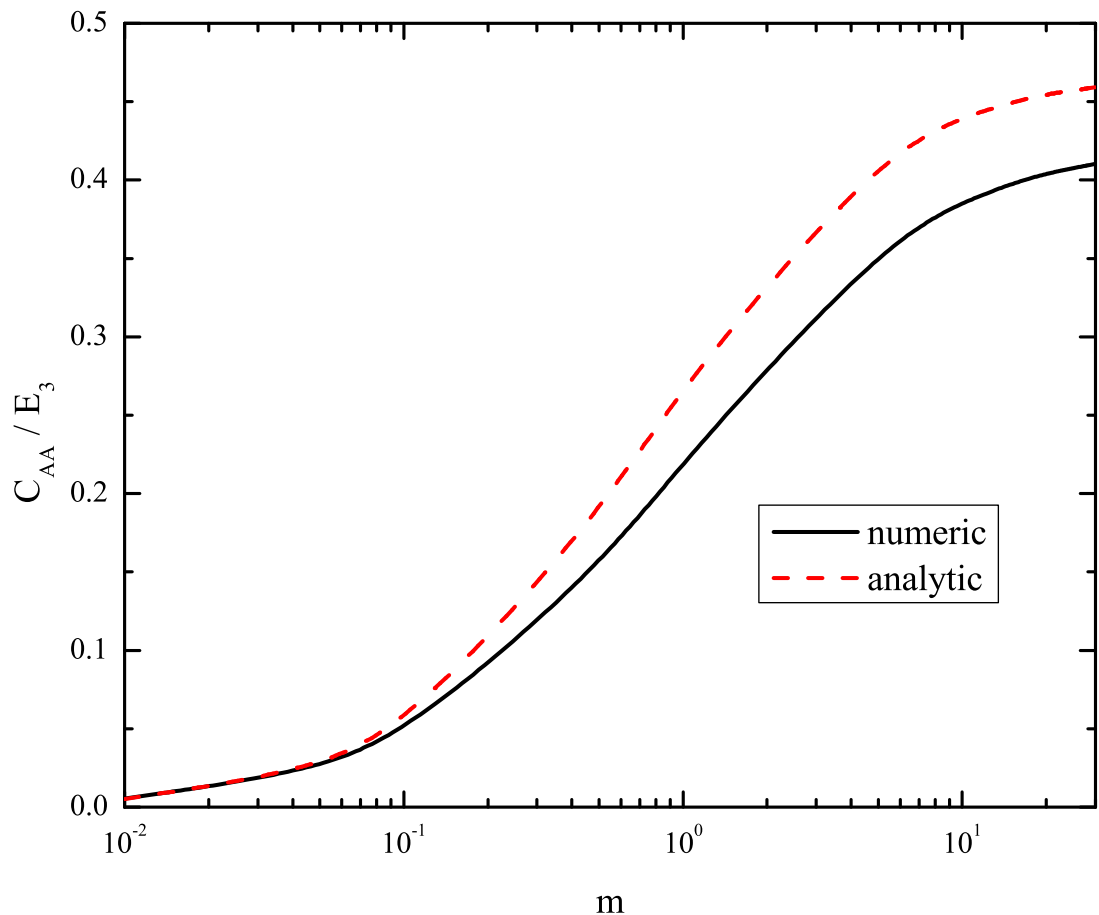}%
\caption{Comparison between the analytic estimative to $C_{AA}$, given in (\ref{eq.11}) and the numerical calculation in (\ref{eq.09a}).}%
\label{fig.01}%
\end{figure}

We can estimate the two-body contact parameter dependence on
excitation energy by use of the approximation to the ground state in
(\ref{eq.A11}). Inserted in (\ref{eq.09a}) we find an
expression for $\frac{C_{AA}}{E_3}$, that is
\begin{equation}
\frac{C_{AA}}{E_3}= 16 \pi \frac{m^2}{(1+m)(2+m)} f_A^2(0) \left(1+ \frac{2}{\ln(E_3)}+\frac{2}{\ln^2(E_3)}\right) .
\label{eq.11}
\end{equation}  
A comparison between this approximation and the numerical results is
shown in Fig. \ref{fig.01}.  We see that (\ref{eq.11}) provides a
fairly good estimate, which is accurate within $5\%$ for small $m$,
around $10\%$ for $m>1$, and within about $20\%$ deviation in the
worst case of $m =1$.  The divergence in (\ref{eq.11}) for $E_3
\rightarrow 1$ means that the two-body contact parameters diverge when
the three-body system approaches this threshold of binding.  This does
not reveal the full energy dependence since the normalization factor,
$f_A^2(0)$, also is state and energy dependent.

%
%

The non-universality of the two-body contact parameters does not
encourage universality investigations of the three-body contact
parameter, which is related to a sub-leading order. However, at least
the system with two non-interacting identical particles turned out to
be universal and may lead to an interesting large-momentum three-body
structure.  As before, by inserting $E_{AA}=0$ in the set of coupled
integral equations (\ref{eq.02}) we find $f_C(q_C)=0$.  Then
(\ref{eq.03a}) to (\ref{eq.03d}) show directly that $n_1(q_c)$
and $n_3(q_C)$ vanish when $f_C(q_C)=0$, leaving only possible
contributions from $n_4(q_C)$ and $n_5(q_C)$.

\begin{figure}[!htb]%
\includegraphics[scale=0.5]{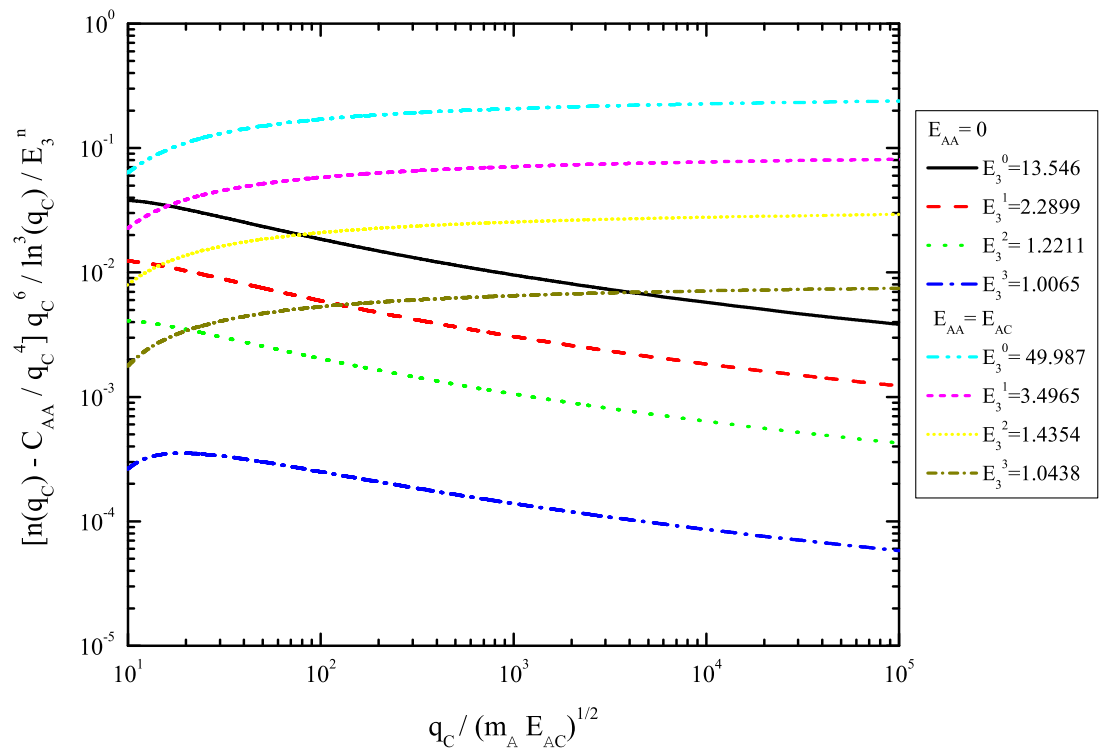}%
\caption{
The sub-leading order of the one-body momentum density divided by $E_3^n$ for each bound state labeled as $n$ in a system composed with two identical ( $A=^{133}$Cs ) particles and a distinct one ( $C=^{6}$Li ) as a function of the momentum $q$ for both $E_{AA}=1$ and $E_{AA}=0$.}%
\label{fig.02}%
\end{figure}

We show in Fig.~\ref{fig.02} the sub-leading order of the
large-momentum distribution multiplied by $q_C^6/\ln^3(q_C)$, that is
$C_{AA,C}$, as functions of $q_C$ for the four bound states for a
system where $A=^{133}$Cs and $C=^{6}$Li and for both $E_{AA}=1$ and
$E_{AA}=0$.  We only show one of these three-body contact parameters 
defined in (\ref{eq.07}) and (\ref{eq.07a})
since the other one, $C_{AC,A}$, is related state-by-state through the
mass factors in (\ref{eq.07}) and (\ref{eq.07a}).  The momentum
dependence approach the predicted constancy at large $q_C$ but by
increasing or decreasing for interacting or non-interacting identical
particles, respectively.  We divided by the three-body energy to see
if a simple energy scaling could explain the differences.  Not
surprisingly, a more complicated and non-universal behavior is
present.

However, it is striking to see that this sub-leading order in the
large-momentum limit is negligibly small for non-interacting compared
to interacting identical particles.  This implies that a negligible
three-body contact parameter combined with a universal two-body
contact parameter can be taken as a signature of a two-body
non-interacting subsystem within a three-body system in 2D.

%

\section{Discussion and Outlook}\label{discussion}
In this work, we have considered three-body systems with attractive 
zero range interactions for general masses and interaction strengths 
in two dimensions using the Faddeev decomposition to write the 
momentum-space wave function, through which the one-body momentum 
density is obtained. The momentum density tail gives the two- and 
three-body contact parameters, namely $C_2$ and $C_3$, respectively. 
We derived analytic expressions for the asymptotic spectator functions 
and for both $C_2$ and $C_3$ for three distinguishable bosons. 

We found that the asymptotic spectator function for each of the three
distinguishable particles has the same functional form as calculated
for three identical particles in \cite{bel13a}. Moreover, we showed
that the three distinct spectator functions relate to each other
through a constant, $\Gamma$, properly weighted by reduced
masses. These analytic results are supported by accurate numerical
calculation, which confirmed both the asymptotic behavior and the
relation between the asymptotic expressions for different spectator
functions in a generic case of three distinguishable particles.  

The spectator functions and their asymptotic behavior define
both two- and three-body contact parameters, $C_2$ and
$C_3$. The parameters $C_2$ arise from integration
of the spectator functions over all momenta, and both small and large
momenta contribute. In the case of the ground state, we are able
to use our knowledge of the asymptotics of the spectator function
to infer the behavior for all momenta.
We found that the three two-body parameters for a
system of three distinguishable particles are related by simple mass
scaling.  However, these two-body contact parameters are in general
not universal in the sense of being independent of the state when more
than one excited state is present.  
In contrast, we find universality for three-body systems with one
distinguishable and two identical, non-interacting particles. In that
case the third particle apparently does not disturb the short-distance
structure arising from the two interacting
particles.  Hence, the two-body contact parameter turn out to be
universal. This is similar to the 3D case and three identical bosons
where $C_2$ is universal in the scaling or Efimov limit where the 
binding energy is negligible \cite{castin2011}.

In 3D systems, the two-body contact has been observed 
in experiments using time-of-flight and mapping to momentum space \cite{stewart2010},
Bragg spectroscopy \cite{stewart2010,kuhnle2010,wild2012,kuhnle2011,hoinka2012}, 
or momentum-resolved photoemission spectroscopy (similar to angle-resolved 
photoemission spectroscopy) \cite{frohlich2012}. Measuring 
the sub-leading term and thus accessing $C_3$ requires more 
precision which has so far only produced upper limits for
the particular case of $^{87}$Rb \cite{wild2012}. In 
2D systems the functional form of the sub-leading term
is different from the 3D case so it is difficult to compare 
the cases. 
However, given that the precision improves continuously it 
should be possible to also probe the 2D case when tightly 
squeezing a 3D sample. As we have shown here, the mass ratio
can change the values of the contact parameters significantly.
We thus expect that mixtures of different atoms is the most
promising direction to make a measurement of a 2D contact
parameter. 

We have analyzed in details two different systems of the heavy-heavy-light 
type that is relevant for current experiments with cold gas mixtures. 
In both cases the light particle is $^{6}$Li while the two heavy particles
are either both $^{133}$Cs or both $^{39}$K.
We find that, unlike the equal mass
scenario, the two-body contact parameters are not universal constants
when all subsystems are interacting. Here universal means that $C_2$
divided by the three-body binding energy is independent of which excited
state is considered. However, if the
two identical particles are not interacting, the heavy-heavy and the 
heavy-light two-body contact
parameters become universal and are related to each other by a factor
of two.  

The methods presented here are in principle also applicable to 1D setups
and it would be interesting to investigate the question of universality
of the contact parameters for three-body states there as well. In some 
respects 1D is easier to handle since zero-range interactions do not
require regularization and one can in fact map the 
3D scattering length to a 1D equivalent \cite{olshanii1998} which 
provides access to confinement-induced resonances that allow the 
study of the infinite 1D coupling strength limit. This was recently 
demonstrated for trapped few-fermion systems in 1D \cite{serwane2011,zurn2012}.
In that case the two-body contact can be determined fully analytically
using the methods described in \cite{lindgren2013,volosniev2013}. It
would be very interesting to consider the bosonic case where three-body 
bound states are possible with or without an in-line trap. In the 
case of quasi-1D setups where the transverse trapping energy is
a relevant scale compared to the binding energies, one needs to 
also take into account the transverse (typically harmonic) degrees
of freedom \cite{mora2004,pricoupenko2011}. Our formalism can be 
adapted to this case as well. Another interesting pursuit would be
to long-range interactions using either heteronuclear molecules or 
atoms with large magnetic dipole moments \cite{laheye2009}, where
the momentum distribution has in fact already been probed in experiments 
\cite{wang2010}. Bound state formation has been predicted in both 
single- \cite{ticknor2011}, bi- \cite{klawunn2010a,baranov2011,zinner2012}, and 
multi-layer systems \cite{wang2006,vol2012,jeremy2012}, as well as in one
or several quasi-1D tubes \cite{sinha2007,bartolo2013,guan2013,klawunn2010b,vol2013}.
The current formalism should be adaptable
to dipolar particles and the contact parameters could subsequently be studied. In particular,
in the limit of small binding energy one may in some cases use effective short-range 
interaction terms to mimic dipolar interactions \cite{vol2013} which makes the implementation
through the Faddeev equations considerably simpler. This is of course also the limit in 
which the contact parameters are most interesting.

\paragraph*{Acknowledgments} This work was partly support by funds
provided by FAPESP (Funda\c c\~ao de Amparo \`a Pesquisa do Estado
de S\~ao Paulo) and CNPq (Conselho Nacional de Desenvolvimento
Cient\'\i fico e Tecnol\'ogico ) of Brazil, and by the Danish 
Agency for Science, Technology, and Innovation.

\end{document}